\documentclass[twoside,12pt]{article}
\usepackage{CJK}
\usepackage{indentfirst}
\usepackage{bm}
\usepackage{epsfig}
\usepackage{indentfirst}
\usepackage{picinpar,graphicx}
\usepackage[namelimits]{amsmath} %数学公式
\usepackage{amssymb}             %数学公式
\usepackage{amsfonts}            %数学字体
\usepackage{mathrsfs} 
\usepackage{booktabs}
\usepackage{slashed}
\usepackage{color}
\usepackage{verbatim}
\usepackage{subfigure}
\usepackage{appendix}

\begin{document}
	
	%\begin{CJK*}{GBK}{song}
		
		%-------------------  First Head  -----------------------------------------
		\thispagestyle{empty} \vspace*{0.8cm}\hbox
		to\textwidth{\vbox{\hfill\huge\sf Commun. Theor. Phys.\hfill}}
		\par\noindent\rule[3mm]{\textwidth}{0.2pt}\hspace*{-\textwidth}\noindent
		\rule[2.5mm]{\textwidth}{0.2pt}
		
		%=================== Text begin here =============================================
		
		\begin{center}
			\LARGE\bf Magnetic correction to the Anomalous Magnetic Moment of Electron 
		\end{center}
		
		\footnotetext{\hspace*{-.45cm}\footnotesize $^*$ linfan19@mails.ucas.ac.cn.}
		\footnotetext{\hspace*{-.45cm}\footnotesize $^\dag$ huangmei@ucas.ac.cn }

		\begin{center}
			\rm Fan Lin$^*$ \ and  \ Mei Huang$^\dag$
		\end{center}
		
		\begin{center}
			\begin{footnotesize} \sl
				School of Nuclear Science and Technology, University of Chinese Academy of Sciences, Beijing 100049, China.
			\end{footnotesize}
		\end{center}
		
		\begin{center}
			\footnotesize (Received XXXX; revised manuscript received XXXX)
			
		\end{center}
		
		\vspace*{2mm}
		
		\begin{center}
			\begin{minipage}{15.5cm}
				\parindent 20pt\footnotesize

				We investigate the leading order correction of anomalous magnetic moment (AMM) to the electron in weak magnetic field and find that the magnetic correction is negative and magnetic field dependent, indicating a magnetic catalysis effect for the electron gas. In the laboratory to measure the $g-2$, the magnitude of the magnetic field $B$ is several $\mathrm{T}$, correspondingly the magnetic correction to the AMM of electron/muon is around $10^{-34}$/$10^{-42}$, therefore the magnetic correction can be safely neglected in current measurement. However, when the magnitude of the magnetic field strength is comparable with the electron mass, the magnetic correction of electron's AMM will become considerable. This general magnetic correction to charged fermion's AMM can be extended to study QCD matter under strong magnetic field.
				
			\end{minipage}
		\end{center}
		
		\begin{center}
			\begin{minipage}{15.5cm}
				\begin{minipage}[t]{2.3cm}{\bf Keywords:}\end{minipage}
				\begin{minipage}[t]{13.1cm}
					Anomalous magnetic moment, magnetic correction, magnetic field.
				\end{minipage}\par\vglue8pt
				
			\end{minipage}
		\end{center}
		
		\section{Introduction}
		Recently, the anomalous magnetic moment of the quark \cite{Ferrer:2009nq,Fayazbakhsh:2014mca,Strickland:2012vu,Mao:2018jdo,Mei:2020jzn,Chaudhuri:2019lbw,Ferrer:2015wca,Chaudhuri:2020lga,Xu:2020yag} attracted much interest in investigating QCD matter under strong magnetic field created through non-central heavy ion collisions \cite{Skokov:2009qp,Deng:2012pc}. It has been known that the magnetic field catalyzes the chiral condensate of spin-0 quark-antiquark pair which carries net magnetic moment and triggers a dynamical anomalous magnetic moment (AMM) of quarks \cite{Chang:2010hb,Ferrer:2008dy,Ferrer:2013noa}. The exact value of quark's AMM under magnetic field is unknown, but very important for magnetized QCD matter. In this work, we will learn some experience by investigating the anomalous magnetic moment (AMM) of electron in magnetic field, which can be calculated from theory and measured in experiment with high precision.
		
		As is well-known that a charged fermion can interact with external magnetic field through  an intrinsic magnetic momentum ($\hbar=1,c=1$)
		\begin{equation*}
			\boldsymbol{\mu}=g\left(\frac{q}{2 m}\right) \boldsymbol{s},
		\end{equation*}
		where $q,m,\boldsymbol{s}$ are charge, mass and spin of the fermion, respectively. For electron, the $\mathrm{Land\acute{e}}$ factor $g_{e}$ is believed exactly equal to $2$ according to Dirac’s relativistic quantum mechanics until a small derivation is observed in an elaborate experiment \cite{Kusch:1948mvb}, the deviation of $g_e$ from 2 is defined as the anomalous magnetic moment $a_{e}=(g_{e}-2)/2$. Meanwhile, the quantum electrodynamics (QED) was in the ascendant and deepened our understanding of the physical vacuum, where the creation and annihilation of virtual particles contributes additional loop-diagram corrections. The leading order correction $a_{e}= \alpha/2\pi$ ($\alpha\simeq 1/137$ is the fine structure constant of QED) calculated by J.Schwinger \cite{Schwinger:1948iu} with renormalizable QED matches the experimental measurement perfectly, powerfully manifesting the validity of quantum field theory. 
		
		In the standard model, the theoretical contribution to $a_{e}$ comes from three types of interactions, electromagnetic, hadronic, and electroweak:
		\begin{equation*}
			a_{e}(\text{theory})=a_{e}(\mathrm{QED})+a_{e}(\text {hadronic})+a_{e}(\text {electroweak}),
		\end{equation*}
		where the QED contribution can be expressed with perturbative expansion:
		\begin{equation*}
			a_{e}(\mathrm{QED})=\sum_{n=1}^{\infty}\alpha^{n} a_{e}^{(2 n)}.
		\end{equation*}
		J.Schwinger's consequence gives $a_{e}^{(2)}=1/2\pi$ and the latest calculation from Aoyama et al. \cite{Aoyama:2007dv,Aoyama:2012wj} has reached up to the tenth-order $a_{e}^{(10)}$ with the help of the automatic code generator numerically. The accuracy of experimental electron's AMM $a_{e}(\text{experiment})$ also enhances continually with measurement in one-electron quantum cyclotron \cite{VanDyck:1987ay,Odom:2006zz,Hanneke:2008tm}, which provides us with a surprising consistency \cite{Aoyama:2012wj,Hanneke:2008tm}
		\begin{equation*}
			a_{e}(\text{experiment})-a_{e}(\text {theory})=-1.06(0.82) \times 10^{-12}.
		\end{equation*}
		In essence, the emergence of electron's AMM is related to slightly broken chiral symmetry because of electron mass $m_{e}$ \cite{Chang:2010hb}, implying any dynamical mechanisms can further break the chiral symmetry always inducing an extra AMM. Naturally, the external magnetic field existing in one-electron quantum cyclotron is worthy of consideration, for the study of massless spinor QED explicitly shows that magnetic field reinforces chiral symmetry breaking by endowing electron with a dynamical mass, famous as magnetic catalysis effect \cite{Miransky:2015ava}, which suggests that the magnetic field could provide electron with an extra correction. Recently, the latest experimental measurement of muon $g_{\mu}-2$ confirms the disagreement between experiment and theory \cite{Muong-2:2021ojo,Muong-2:2006rrc}, for the potential possibility as a window to pry into new physics, attracting much attention.  All of these motivate us to investigate electron's AMM in magnetic field and extend to general fermions, which can deepen our understanding of quantum field theory in magnetic field.
		
		In this work, we investigate the leading order correction of magnetic filed on the AMM of electron, which can be calculated with relative high precision. To our knowledge, this has not been done in any other literature. The motivation of this work is two sides: firstly, we would like to check how the magnetic filed will modify the AMM of electrons, secondly, the magnetic filed dependent of electron's AMM will shed light on the AMM of quarks under magnetic field.

		\section{The leading order correction of electron-photon vertex}
		
		The electron-photon coupling in QED is described by $i e \bar{\psi} \gamma^{\mu} \psi A_{\mu} $, virtual particles loops provide the vertex with corrections order by order shown below in Feynman diagrams:
		
		\begin{comment}
			\begin{equation*}
				\feynmandiagram [baseline=(b.base),horizontal=a to b] {
					a -- [photon,momentum=\(q\)] b [blob],
					c -- [fermion,edge label= \(p\)] b -- [fermion,edge label=\( p^{'}\)] d,
				};
				=\quad
				\feynmandiagram [baseline=(b.base),horizontal=a to b] {
					a -- [photon,momentum=\(q\)] b,
					d -- [fermion,edge label= \(p\)] b -- [fermion,edge label=\( p^{'}\)] c,
				};
				+\quad
				\feynmandiagram [small,baseline=(b.base),horizontal=a to b] {
					a -- [photon,momentum=\(q\)] b,
					c --[fermion,edge label= \(p\)] g--[fermion] b -- [fermion] h--[fermion,edge label= \(p^{'}\)] d,
					g --[boson] h,
				};
				+\cdots,
			\end{equation*}
		\end{comment}

\begin{figure}
	    	\centering
	    	\includegraphics[width=0.3\linewidth]{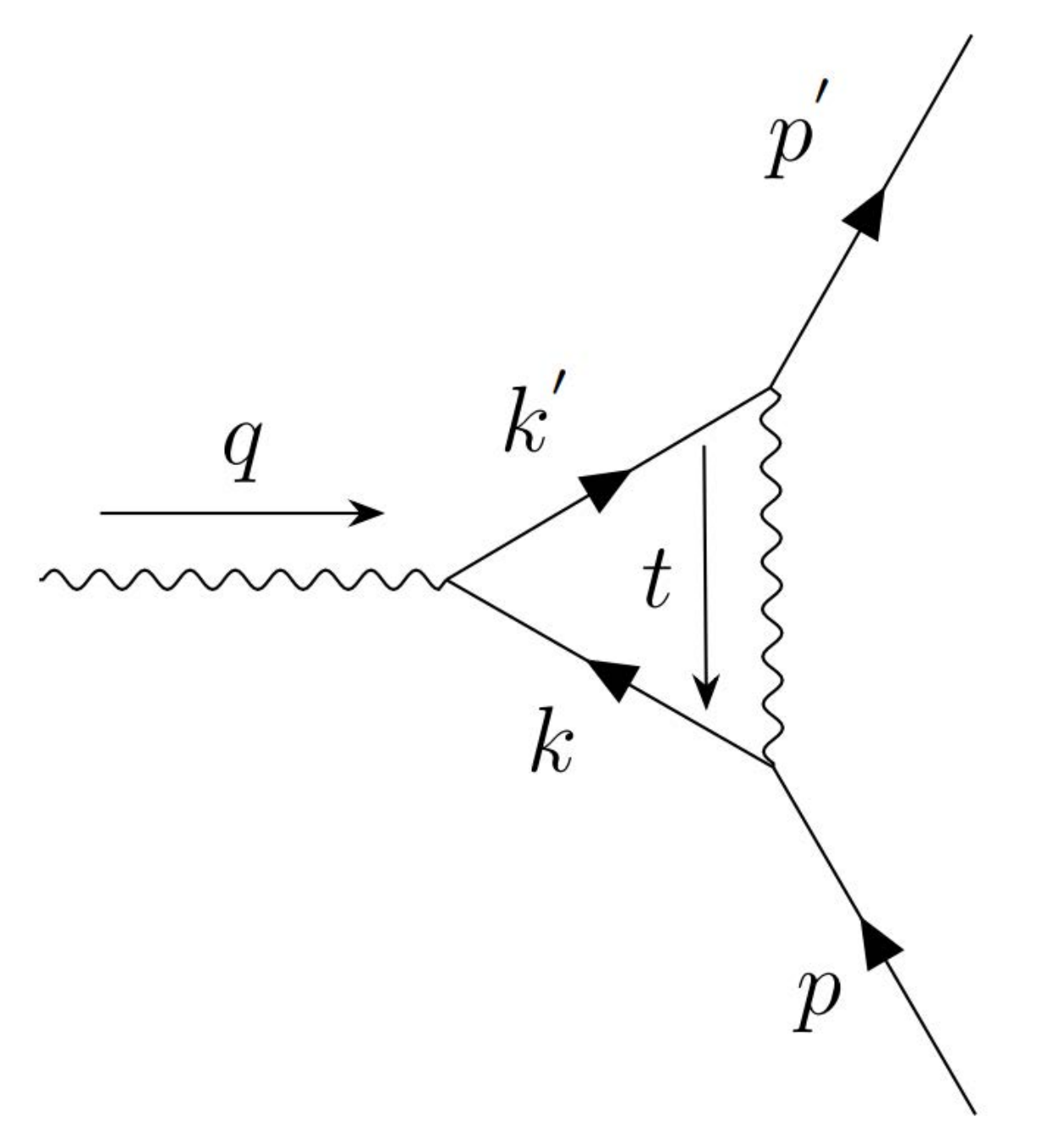}
	    	\caption{Leading order correction of electron-photon vertex. }
	    	\label{leading order}
	    \end{figure}

	 \begin{center}
		\includegraphics[width=0.7\linewidth]{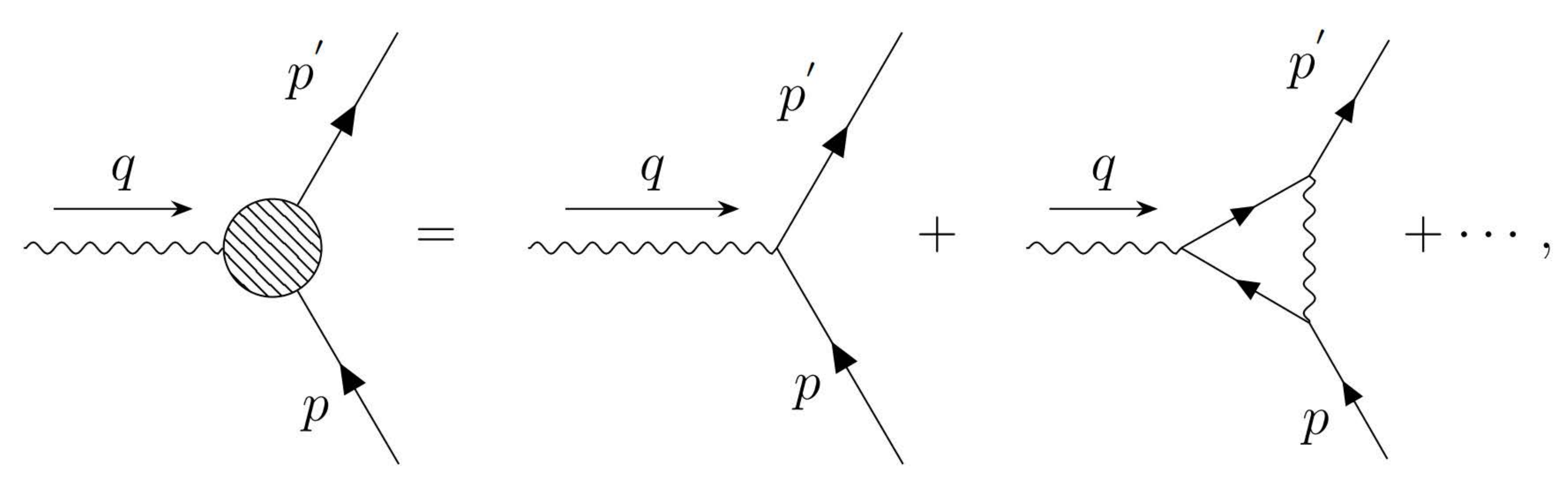}
	  \end{center}

		\noindent where $q=p^{'}-p$ is the momentum transformation and not on-shell $q^{2} \neq 0$, the electromagnetic current changes: $\bar{\psi}\gamma^{\mu}\psi \; \longrightarrow \; \bar{\psi}\Gamma^{\mu}\psi $. A complete set of Dirac matrices contains $\mathbf{1}_{4\times 4},\gamma^{\mu},\sigma^{\mu\nu}\!=\!\frac{\mathrm{i}}{2}[\gamma^{\mu},\gamma^{\nu}],\gamma^{\mu}\gamma^{5},\gamma^{5}$, the expansion of $\Gamma^{\mu}$ in this basis has 16 components. However, the parity symmetry of QED, initial and final electron on-shell, electromagnetic current and momentum conservation excludes all other terms leaving  $\Gamma^{\mu}$  in momentum space taking the following form \cite{Peskin:1995ev}:
		\begin{equation}
			\Gamma^{\mu}=\gamma^{\mu}F_{1}(q^{2})+\frac{\mathrm{i} \sigma^{\mu\nu} q_{\nu}}{2m} F_{2}(q^{2}),
			\label{total vertex}
		\end{equation}
		where $m$ is the electron mass and $F_{1}, F_{2}$ are form factors and $ a_{e}(\mathrm{QED})=F_{2}(0)$ \cite{Peskin:1995ev}. A point-particle in the absence of radiative
		corrections has $F_{1} \equiv 1,F_{2} \equiv 0$ which derives the Dirac’s value for the magnetic moment. Schwinger's calculation of leading order correction shown in Fig.(\ref{leading order}) gives $F_{2}(0)=\alpha/2\pi$, which contribute the biggest correction. 
		
		More carefully, AMM term $\bar{u}(p^{'})\sigma^{\mu\nu} u(p)$ denotes a helicity-flipping interaction not invariant under chiral transformation $\psi(x) \rightarrow \exp \left(i \theta \gamma^{5}\right) \psi(x)$, which indicates the AMM effect emerges from chiral symmetry breaking. Actually, Gordon's identity is derived from Dirac's equation:
		\begin{equation}
			2 m \bar{u}(p^{'}) \gamma^{\mu} u\left(p\right)=\bar{u}(p^{'})\left[2 (p^{'}+p)^{\mu}+i \sigma^{\mu \nu} q_{\nu}\right] u\left(p\right),
		\end{equation}
		this formula explicitly exhibits mass term causing left-handed fermion entangles with right-handed one. If we naively set $m=0$, it leads to the massless spinor QED, which is an infrared-free theory, where photon and electron are not coupled, then there will be no observable AMM, i.e., $F_{2}\equiv 0$. Nevertheless, this does not conflict with the perturbative and mass-independent consequence $F_{2}(0)=\alpha/2\pi$ in one-loop calculation, because QED is not an asymptotically free theory and does not possess a well-defined chiral limit, so the perturbative expression and computation of $F_{2}(0)$ cannot apply to the case $ m= 0$. On the contrary, the quantum chromodynamics (QCD) is asymptotically free whose chiral limit could be rigorously defined nonperturbatively, so quark's AMM as a continuous function in chiral limit is expected. This has been confirmed by Dyson-Schwinger equation calculation \cite{Chang:2010hb}, which reveals a dressed-quark with mass $M_{q}\sim 0.5 \,\mathrm{GeV}$ possessing an AMM $\kappa_{q}\propto M_{q}$. In addition, $\kappa_{q} \sim 0.1 $ is much bigger than an intrinsic one due to the strong breaking of QCD's chiral symmetry, dramatically changing the nature of quark matter in magnetic field \cite{Xu:2020yag}. The presence of magnetic field induces magnetic catalysis effect \cite{Shovkovy:2012zn},  but it is unknown how this would affect the AMM of quarks. It is quite challenging to do the direct calculation in QCD because of the nonperturbative nature, therefore we will check the magnetic field dependent behavior of electron's AMM and use it for reference in the case of quarks.
		
		\par To calculate the magnetic field correction to the  AMM of electron, we put the Lagrangian density of QED in an external magnetic field, 
		\begin{eqnarray}
			\mathcal{L}=\bar{\psi}\left(i \gamma^{\mu} D_{\mu}-m \right) \psi-\frac{1}{4} F^{\mu \nu} F_{\mu \nu}
		\end{eqnarray}
		where the covariant derivative $D_{\mu}=\partial_{\mu}-i e\left(A_{\mu}^{\mathrm{e x t}}+A_{\mu}\right)$, the vector potential $A_{\mu}^{\mathrm{e x t}}$ introduce an external and uniform magnetic field $\mathbf{B}=(0,0,B)$ along $z$ direction, we choose symmetric gauge $A_{\mu}^{\mathrm{e x t}}=\frac{B}{2}(0,y,-x,0)$ for convenience in the later calculations. The presence of a magnetic field breaks translation invariance, the amplitude of the process has to be computed in coordinate space and subsequently integrated over space-time \cite{Ayala:2017vex}. The photon does not carry charge so the propagator keeps:
		\begin{equation}
			G_{\mu\nu}(p)=\frac{-ig_{\mu\nu}}{p^2+i\epsilon}.
		\end{equation}
		For the electron propagator under a uniform magnetic field, the proper-time approach is adopted in coordinate space representation \cite{Schwinger:1951nm}:
		\begin{equation}
			S\left(x, x^{\prime}\right)=\Phi\left(x, x^{\prime}\right) \int \frac{\mathrm{d}^{4} k}{(2 \pi)^{4}} e^{-i k \cdot\left(x-x^{\prime}\right)} \tilde{S}(k),
		\end{equation}
		where 
		\begin{equation}
			\Phi\left(x, x^{\prime}\right)=\exp \left\{i\left|q \right| \int_{x^{\prime}}^{x} \mathrm{d} \xi^{\mu}\left[A_{\mu}+\frac{1}{2} F_{\mu \nu}\left(\xi-x^{\prime}\right)^{\nu}\right]\right\}
			\label{Schwinger phase}
		\end{equation}
		is the Schwinger's phase factor, the Fourier transformation of translation invariant part $\tilde{S}(k)$ is given as: 
		\begin{equation}
			\begin{aligned}
				\tilde{S}(k)=& \int_{0}^{\infty} \mathrm{d} s \, e^{i s\left(k_{\|}^{2}-m^{2}-k_{\perp}^{2} \frac{\tan \left(|q B|s\right)}{|qB|s}+i \epsilon\right)}\left[\slashed{k}+m+\left(k^{1} \gamma^{2}-k^{2} \gamma^{1}\right) \tan (|qB|s)\right]\left[1-\gamma^{1} \gamma^{2} \tan (|qB|s)\right],
			\end{aligned}
		\end{equation}
		here $s$ is Schwinger's proper-time, the basic notations are defined as:
		\begin{equation*}
			k_{\perp}^{\mu}\equiv(0,k^{1},k^{2},0),\quad k_{\|}^{\mu}\equiv (k^{0},0,0,k^{3}) ,\quad 
			k_{\perp}^{2}\equiv k_{1}^{2}+k_{2}^{2},\quad k_{\|}^{2}\equiv k_{0}^{2}-k_{3}^{2},\quad  k^{2}= k_{\|}^{2}-k_{\perp}^{2}.
		\end{equation*}
		here $q$ represents the charge of the fermion and we assume $ eB > 0 $ for convenience.

		\begin{comment}
			\begin{figure}
				\centering
				\feynmandiagram [baseline=(b.base),horizontal=a to b] {
					a -- [photon,momentum=\(q\)] b,
					c --[fermion,edge label= \(p\)] g--[fermion,edge label= \(k\)] b -- [fermion,edge label= \(k^{'}\)] h--[fermion,edge label= \(p^{'}\)] d,
					g --[photon, rmomentum = \(t\)] h,
				};
				\caption{Leading order correction of electron-photon vertex. }
				\label{leading order}
			\end{figure}
		\end{comment}
	
		\par By using perturbative method, we can expand the total vertex amplitude with respect to $\alpha$ in leading-order
		% \par The formal expression of the leading order vertex correction amplitude is 
		\begin{equation}
			\bar{u}(p^{'})\Gamma^{\mu}u(p)=\bar{u}(p^{'})\gamma^{\mu}u(p)+\bar{u}(p^{'})\mathcal{M}^{\mu}u(p)+\mathcal{O}(\alpha).
		\end{equation}
		As is shown in Figure \ref{leading order}, the leading-order correction $\mathcal{M}^{\mu}$ reads:
		\begin{equation}
			\begin{aligned}
				\mathcal{M}^{\mu} &=\int \mathrm{d}^{4} x \mathrm{d}^{4} y \mathrm{d}^{4} z \, \Phi(x, y) \Phi(y, z) \int \frac{\mathrm{d}^{4} t}{(2 \pi)^{4}} \frac{\mathrm{d}^{4} k^{'}}{(2 \pi)^{4}} \frac{\mathrm{d}^{4} k}{(2 \pi)^{4}}\\
				& \times e^{-i t \cdot(z-x)} e^{-i k \cdot(x-y)} e^{-i k^{'} \cdot(y-z)} e^{i p \cdot x} e^{i q \cdot y} e^{-i p^{'}\cdot z}  G_{\nu\rho}(t)(-ie\gamma^{\nu})\tilde{S}(k^{'})\gamma^{\mu}\tilde{S}(k)(-ie\gamma^{\rho}),\\
				& =\int \frac{\mathrm{d}^{4} t}{(2 \pi)^{4}} \frac{\mathrm{d}^{4} k^{'}}{(2 \pi)^{4}} \frac{\mathrm{d}^{4} k}{(2 \pi)^{4}}
				\times \frac{i e^2}{t^2} \int_{0}^{\infty} \mathrm{d}r \int_{0}^{\infty} \mathrm{d}s 
				e^{i r\left(k_{\|}^{'2}-m^{2}-k_{\perp}^{'2} \frac{\tan \left(e Br\right)}{eBr}\right)}
				e^{i s\left(k_{\|}^{2}-m^{2}-k_{\perp}^{2} \frac{\tan \left(e Br\right)}{eBs}\right)}\\
				& \times \gamma^{\nu}\left[\slashed{k}^{'}+m+\left(k^{'1} \gamma^{2}-k^{'2} \gamma^{1}\right) \tan (eBr)\right]\left[1-\gamma^{1} \gamma^{2} \tan (eBr)\right] \gamma^{\mu}\\
				&\times \left[\slashed{k}+m+\left(k^{1} \gamma^{2}-k^{2} \gamma^{1}\right) \tan (eBs)\right]\left[1-\gamma^{1} \gamma^{2} \tan (eBs)\right]\gamma_{\nu}\\
				& \times \int \mathrm{d}^{4} x \mathrm{d}^{4} y \mathrm{d}^{4} z \, \Phi(x, y) \Phi(y, z) e^{-i t \cdot(z-x)} e^{-i k \cdot(x-y)} e^{-i k^{'} \cdot(y-z)} e^{i p \cdot x} e^{i q \cdot y} e^{-i p^{'}\cdot z},
			\end{aligned}
		\end{equation}
		where $x,y,z$ is three space-time vertexes with corresponding momentum, $ s,r$ and $\Phi(x, y), \Phi(y, z)$ are proper-time and phase factors emerging from two electron propagators.  The expression of $\mathcal{M}^{\mu}$ is quite tedious, in the following we separate $\mathcal{M}^{\mu}$ into three parts: 1) the phase factor $P(t,k,k^{'})$, 2) the momentum integrand $\mathcal{A}^{\mu}(t,k,k^{'})$, and 3) the $\gamma$ matrix part $\Theta^{\mu}(r,s)$ with their explicit expressions given below:
		\begin{equation}
			P(t,k,k^{'})=\int \mathrm{d}^{4} x \mathrm{d}^{4} y \mathrm{d}^{4} z \, \Phi(x, y) \Phi(y, z) e^{-i t \cdot(z-x)} e^{-i k \cdot(x-y)} e^{-i k^{'} \cdot(y-z)} e^{i p \cdot x} e^{i q \cdot y} e^{-i p^{'}\cdot z},
		\end{equation}
		\begin{equation}
			\begin{aligned}
				\mathcal{A}^{\mu}(t,k,k^{'})&=G_{\nu\rho}(t)(-ie\gamma^{\nu})\tilde{S}(k^{'})\gamma^{\mu}\tilde{S}(k)(-ie\gamma^{\rho})=\frac{i e^2}{t^2} \int_{0}^{\infty} \mathrm{d}r \int_{0}^{\infty} \mathrm{d}s \\
				&  \times e^{i r\left(k_{\|}^{'2}-m^{2}-k_{\perp}^{'2} \frac{\tan \left(e Br\right)}{eBr}\right)}
				e^{i s\left(k_{\|}^{2}-m^{2}-k_{\perp}^{2} \frac{\tan \left(e Br\right)}{eBs}\right)}\times \Theta^{\mu}(r,s),
			\end{aligned}
		\end{equation}
		\begin{equation}
			\begin{aligned}
				\Theta^{\mu}(r,s)&=\gamma^{\nu}\left[\slashed{k}^{'}+m+\left(k^{'1} \gamma^{2}-k^{'2} \gamma^{1}\right) \tan (eBr)\right]\left[1-\gamma^{1} \gamma^{2} \tan (eBr)\right] \gamma^{\mu}\\
				&\times \left[\slashed{k}+m+\left(k^{1} \gamma^{2}-k^{2} \gamma^{1}\right) \tan (eBs)\right]\left[1-\gamma^{1} \gamma^{2} \tan (eBs)\right]\gamma_{\nu}.
			\end{aligned}
			\label{Theta}
		\end{equation}
		In the next subsections, we will carefully handle these three parts, respectively.

		\subsection{The phase factor part}
		In this section, we deal with the Schwinger part $P(t,k,k^{'})$. With the symmetric gauge $A^{\mu}=\frac{B}{2}(0,-y,x,0)$ and Eq.(\ref{Schwinger phase}),  the only non-zero terms of $F_{\mu\nu}$ are $ F_{12}=-F_{21}=-B $ , so the product of the phase factor can be written as
		\begin{equation}
			\Phi(x, y) \Phi(y, z)=e^{i\cdot \frac{eB}{2} (\epsilon_{ij} x^{j} y^{i}+\epsilon_{ij} y^{j} z^{i})}  = e^{i\cdot \frac{eB}{2} \epsilon_{ij} y^{i} (x-z)^{j}}.
		\end{equation}
		Introducing a variable $h=x-z$, one has $x=h+z, \mathrm{d}^{4} x=\mathrm{d}^{4} h$  and after the integration of the space-time, we have:
		\begin{equation}
			\begin{aligned}
				P(t,k,k^{'})&=\int \mathrm{d}^{4} x \mathrm{d}^{4} y \mathrm{d}^{4} z \, \Phi(x, y) \Phi(y, z) e^{-i t \cdot(z-x)} e^{-i k \cdot(x-y)} e^{-i k^{'} \cdot(y-z)} e^{i p \cdot x} e^{i q \cdot y} e^{-i p^{'}\cdot z}\\
				&=(2\pi)^{2} 4l^{4} e^{-i\cdot2l^{2}(p+t-k)^{1}(q+k-k^{'})_{2}} e^{i\cdot 2l^{2}(p+t-k)^{2}(q+k-k^{'})_{1}}\\
				& \times (2\pi)^{4}\delta^{4}(p-k+k^{'}-p^{'})(2\pi)^{2}\delta^{2}(p_{\|}+t_{\|}-k_{\|})(2\pi)^{2}\delta^{2}(q_{\|}+k_{\|}-k^{'}_{\|}),
			\end{aligned}
			\label{phase factor}
		\end{equation}
		where $l=1/\sqrt{|eB|}$ is the magnetic length. It is worthy of mentioning that the longitudinal momentum conserves strictly but the transverse momentum expresses a distribution function. The magnetic field causes dimension reduction leaving the longitudinal physics unchanged but the transverse remains a little  subtle. Now $\mathcal{M}^{\mu}$ becomes:
		\begin{equation}
			\begin{aligned}
				\mathcal{M}^{\mu} & = \int (\mathrm{d}t_{\perp} \mathrm{d}k^{'}_{\perp}\mathrm{d}k) (2\pi)^{2}\delta^{2}(p^{'}_{\|}-p_{\|}=q_{\|}) (2\pi)^{2}\delta^{2}(p_{\perp}-k_{\perp}=p^{'}_{\perp}-k^{'}_{\perp}) \\
				& \times (2\pi)^{2} 4l^{4} e^{-i\cdot 2l^{2}(p+t-k)^{1}(q+k-k^{'})_{2}} e^{i\cdot 2l^{2}(p+t-k)^{2}(q+k-k^{'})_{1}} \times \mathcal{A}^{\mu}(t,k,k^{'})\\
				& = \Delta \int (dt_{\perp} dk^{'}_{\perp}dk) (2\pi)^{2} 4l^{4} e^{-i\cdot 2l^{2}(t-k)^{1}(k-k^{'})_{2}} e^{i\cdot 2l^{2}(t-k)^{2}(k-k^{'})_{1}} \times \mathcal{A}^{\mu}(t,k,k^{'}),
			\end{aligned}
			\label{pq0appro}
		\end{equation}
		where
		\begin{equation}
			\Delta=(2\pi)^{2}\delta^{2}(p^{'}_{\|}-p_{\|}=q_{\|})(2\pi)^{2}\delta^{2}(p_{\perp}-k_{\perp}=p^{'}_{\perp}-k^{'}_{\perp})
		\end{equation}
		with $(\mathrm{d}k)\equiv\frac{\mathrm{d}^{4} k}{(2 \pi)^{4}},(\mathrm{d}k_{\perp })\equiv\frac{\mathrm{d}^{2} k_{\perp}}{(2 \pi)^{2}},(\mathrm{d}k_{\| })\equiv\frac{\mathrm{d}^{2} k_{\|}}{(2 \pi)^{2}}$ for simplification, and variable of $\delta$ functions are identities for clarity. The AMM is defined when $q$ is very small and the measurement of AMM proceeds in a low-energy condition, which suggests a reasonable approximation: $q\rightarrow 0,p\rightarrow 0$ applied in last line of Eq.(\ref{pq0appro}), so the momentum conservation conditions: $t_{\|}=k_{\|}-p_{\|},k^{'}_{\|}=k_{\|}+q_{\|}$ becomes $t_{\|}=k_{\|},k^{'}_{\|}=k_{\|}$.

		\subsection{The momentum integrand part}
		The amplitude except the phase factor in the momentum space takes the form of: 
		\begin{equation}
			\begin{aligned}
				\mathcal{A}^{\mu}(t,k,k^{'})&=\frac{i e^2}{t^2} \int_{0}^{\infty} \mathrm{d}r \int_{0}^{\infty} \mathrm{d}s
				\, e^{i r\left(k_{\|}^{'2}-m^{2}-k_{\perp}^{'2} \frac{\tan \left(e Br\right)}{eBr}\right)}
				e^{i s\left(k_{\|}^{2}-m^{2}-k_{\perp}^{2} \frac{\tan \left(e Br\right)}{eBs}\right)}\times \Theta^{\mu}(r,s).
			\end{aligned}
		\end{equation}
		The AMM part can be extracted from $\Theta^{\mu}(r,s)$ later and irrelevant to the momenta $t,k,k^{'}$, so we can integrate all three momenta:
		\begin{equation}
			\begin{aligned}
				\mathcal{I} &=(2\pi)^{2} 4l^{4} \int_{0}^{\infty} \mathrm{d}r \int_{0}^{\infty} \mathrm{d}s  \, \int (\mathrm{d}t_{\perp} \mathrm{d}k^{'}_{\perp}\mathrm{d}k) \frac{i e^2}{k^{2}_{\|}-t^{2}_{\perp}} \\
				&\times e^{-i\cdot 2l^{2}(t-k)^{1}(k-k^{'})_{2}} e^{i\cdot 2l^{2}(t-k)^{2}(k-k^{'})_{1}} e^{i r\left(k_{\|}^{2}-m^{2}-k_{\perp}^{'2} \frac{\tan \left(e Br\right)}{eBr}\right)} e^{i s\left(k_{\|}^{2}-m^{2}-k_{\perp}^{2} \frac{\tan \left(eBr\right)}{eBs}\right)}
				,	\end{aligned}
		\end{equation}
		where $t^{2}=k^{2}_{\|}-t^{2}_{\perp}$ by momentum conservation. In order to remove the pole of momentum integral, We do a Wick rotation transferring Minkowski space-time into Euclidean one \cite{Miransky:2015ava}: $k^{0} \rightarrow ik^{0},r \rightarrow -ir,s \rightarrow -is$. So the integral become 
		\begin{equation}
			\begin{aligned}
				\mathcal{I} &=(2\pi)^{2} 4l^{4} \times (-i)^{2} \int_{0}^{\infty} dr \int_{0}^{\infty} ds  \,\times (i) \int (dt_{\perp} dk^{'}_{\perp}dk_{E}) \frac{-i e^2}{k^{2}_{\|E}+t^{2}_{\perp}}  \\
				& \times e^{-i\cdot 2l^{2}(t-k)^{1}(k-k^{'})_{2}} e^{i\cdot 2l^{2}(t-k)^{2}(k-k^{'})_{1}}  e^{-r\left(k_{\|E}^{2}+m^{2}\right)-k_{\perp}^{'2} l^{2} \tanh \left(e Br\right)}  e^{-s\left(k_{\|E}^{2}+m^{2}\right)-k_{\perp}^{2} l^{2} \tanh \left(e Bs\right)},
			\end{aligned}
			\label{integralI}
		\end{equation}
		here, we use $\tan(-ix)=-i\tanh(x),k_{\|E}=k^{2}_{0}+k^{2}_{3}$, the integral about proper-time $r $ or $s$, momentum $k^{0}$ and photon propagator contribute $-i,i,-1$ under Wick rotation, respectively. Even in Euclidean space, the integral calculation is  still tedious and complicated, we put all details of calculation in Appendix B. Finally, we obtain a much transparent expression:
		\begin{equation}
			\mathcal{I}  = -\frac{\alpha}{4\pi l^{2}}\int_{0}^{\infty} \int_{0}^{\infty}\mathrm{d}r \mathrm{d}s \frac{  e^{-(r+s)m^{2}}}{\left[\tanh(eBr)+\tanh(eBs)\right]}  \frac{\ln v}{v-1}
			\label{mathcalI}
		\end{equation}
		with fine structure constant $\alpha = e^{2}/4\pi$ and  $v=(eBr+eBs)/\tanh(eBr+eBs)$ .
		
		\subsection{The AMM term extracted from  $\Theta^{\mu}(r,s)$}
		
		The presence of the magnetic field breaks the original parity symmetry, thus the leading-order vertex structure Eq.(\ref{Theta}) becomes more complicated than the original one in Eq.(\ref{total vertex}). Fortunately, we just need to extract and calculate the term related to the AMM and not to get stuck in the swamp of total $\Theta^{\mu}(r,s)$. As is seen in Eq.(\ref{total vertex}), the linear term of $q$ corresponds to the AMM effect in absence of magnetic field. We naturally have the momentum constraint in the longitudinal direction $q_{\|}=k^{'}_{\|}-k_{\|}$ which helps to separate the linear term of $q$ but the momentum constraint in the transverse direction emerges as a distribution function Eq.(\ref{phase factor}). Therefore, it is impossible to have the same definition of AMM as in the vacuum, but considering the phase factor Eq.(\ref{phase factor}) is a highly oscillating function in weak magnetic field region which indicates that the momentum $q_{\perp} = k^{'}_{\perp}-k_{\perp}$ substantially dominates the distribution. So under the approximation  $ q_{\perp} \simeq k^{'}_{\perp}-k_{\perp}$, the linear part of $q$ in $\Theta^{\mu}$ can be subtracted and takes the following form:
		\begin{equation}
			\begin{aligned}
				\Theta^{\mu}(r,s)_{q- \mathrm{linear} }&=\gamma^{\nu}\slashed{\tilde{q}}\left[1-\gamma^{1} \gamma^{2} \tan (eBr)\right]\gamma^{\mu}\\
				&\left[\slashed{k}+m+\left(k^{1} \gamma^{2}-k^{2} \gamma^{1}\right) \tan (eBs)\right]\left[1-\gamma^{1} \gamma^{2} \tan (eBs)\right]\gamma_{\nu},
			\end{aligned}
		\end{equation}
		where $\tilde{q}=q+\hat{q},\hat{q}:=\left(0,q_{2}\tan(eBr),-q_{1}\tan(eBr),0\right)$. It is noticed that only the terms consisting of the product of even $\gamma$ matrices contribute to the AMM, thus we directly get rid of other terms and obtain:
		\begin{equation}
			\begin{aligned}
				\Theta^{\mu}&(r,s)_{\mathrm{AMM}}=\gamma^{\nu}\slashed{\tilde{q}}\left[1-\gamma^{1} \gamma^{2} \tan (eBr)\right]\gamma^{\mu} m \left[1-\gamma^{1} \gamma^{2} \tan (eBs)\right]\gamma_{\nu}\\
				&= m\gamma^{\nu}\slashed{\tilde{q}}  \left[ \gamma^{\mu}-\gamma^{\mu}\gamma^{1}\gamma^{2}\tan(eBs)-\gamma^{1}\gamma^{2}\gamma^{\mu}\tan(eBr)+\gamma^{1}\gamma^{2}\gamma^{\mu}\gamma^{1}\gamma^{2}\tan(eBr)\tan(eBs) \right] \gamma_{\nu}.
			\end{aligned}
		\end{equation}
		The presence of magnetic field creates anisotropy between $x\!-\!y$ plane and $z$ direction so that the AMM correction becomes different in the longitudinal and transverse directions. While a magnetic field along $z$ axis is imposing, the classical interaction term $- \boldsymbol{\mu} \cdot \mathbf{B}$, $\boldsymbol{\mu}=(\mu_{x},\mu_{y},\mu_{z})$ implies only $\mu_{z}$ is measurable, corresponding to $\mu=1,2$ components of the magnetic vector potential $A^{\mu}$. Thus the magnetic interaction amplitude is given by 
		\begin{equation}
			\bar{u}(p^{'})\mathcal{M}^{i}u(p)\tilde{A}^{i}=\bar{u}(p^{'})\mathcal{M}^{1}u(p)\tilde{A}^{1}+\bar{u}(p^{'})\mathcal{M}^{2}u(p)\tilde{A}^{2}
			\label{MAcoupling}
		\end{equation}
		where $\tilde{A}^{\mu}$ is the Fourier transformation of $A^{\mu}$, which guides us to concentrate on $\Theta^{\mu} $ with $\mu=1,2$ components. Because two transverse directions share the same physics, therefore we just need to display the case $\mu=1$.
		\par When $\mu=1$, we can simplify the expression into a more friendly form by the Dirac's  algebra: 
		\begin{equation}
			\begin{aligned}
				\Theta^{\mu}(r,s)_{\mathrm{AMM}}&=\frac{i\sigma^{1\alpha} q_{\alpha}}{2m}\{-8m^{2}\left[1+\tan(eBr)\tan(eBs)\right]\} +\frac{i\sigma^{2\alpha} q_{\alpha}}{2m}\{-8m^{2}\left[\tan(eBs)-\tan(eBr)\right]\}\\
				&+\frac{i\sigma^{1\alpha}\hat{q}_{\alpha}}{2m}\{-8m^{2}\left[1+\tan(eBr)\tan(eBs)\right]\} +\frac{i\sigma^{2\alpha}\hat{q}_{\alpha}}{2m}\{-8m^{2}\left[\tan(eBs)-\tan(eBr)\right]\}\\
				&+ 4 m \slashed{q}\left[\gamma^{1}+\gamma^{2}\tan(eBs)-\gamma^{2}\tan(eBr)+\gamma^{1}\tan(eBr)\tan(eBs)\right]\\
				&+4m\slashed{\hat{q}}\left[\gamma^{1}+\gamma^{2}\tan(eBs)-\gamma^{2}\tan(eBr)+\gamma^{1}\tan(eBr)\tan(eBs)\right].
			\end{aligned}
			\label{Thetamu1}
		\end{equation}
		The presence of the magnetic field breaks the concise structure of electron-photon vertex Eq.(\ref{total vertex}) in the vacuum. From the above complicated expression, it's difficult to read explicit information about the magnetic correction of AMM term. 
		
		Obviously, the first term possesses the standard form contributing magnetic correction of AMM, the other terms are much veiled whose physical meaning will be explained by seeking corresponding non-relativistic limit as analyzed in Appendix C.  Finally, the last line also contributes corrections to AMM. Extracting the coefficient of total magnetic correction term of the electron in Eq.(\ref{Thetamu1}) gives 
		\begin{equation}
			8m^{2}\left[\tan^{2}(eBr)-2\tan(eBr)\tan(eBs)\right]
		\end{equation} 
		Some non-invariant terms under gauge transformation that cannot present observable phenomenon are neglected.
		
		\section{Conclusion and discussion}
		For a measurable AMM correction, we only need to calculate the major contribution of $\mathcal{M}^{\mu}$ in $x,y$ direction. Extracting relevant term from Eq.(\ref{Thetamu1}) and do a Wick rotation $r \rightarrow -ir , s \rightarrow -is $,
		\begin{equation}
			8m^{2}\left[ \tan^{2}(eBr)-2\tan(eBr)\tan(eBs)\right] \  \longrightarrow \ 8m^{2}\left[2\tanh(eBr)\tanh(eBs)- \tanh^{2}(eBr)\right]  .
		\end{equation}
		Therefore, one combines meaningful part with integral $\mathcal{I}$ in Eq.(\ref{mathcalI})  
		and obtains the magnetic correction to the electron: 
		\begin{equation}
			\begin{aligned}
				\kappa_B & = -\frac{\alpha }{4\pi l^{2}}\int_{0}^{\infty} \int_{0}^{\infty}\frac{  e^{-(r+s)m^{2}}  \mathrm{d}r\mathrm{d}s }{\left[\tanh(eBr)+\tanh(eBs)\right]}  \frac{\ln v}{v-1}  \times 8m^{2} \left[2\tanh(eBr)\tanh(eBs)- \tanh^{2}(eBr)\right]  \\
				& =-\frac{4m^{2} eB\alpha}{2\pi}     \int_{0}^{\infty} \int_{0}^{\infty}e^{-(r+s)m^{2}}\frac{\left[2\tanh(eBr)\tanh(eBs)- \tanh^{2}(eBr)\right] }  {\left[\tanh(eBr)+\tanh(eBs)\right]}  \frac{\ln v}{v-1}  \mathrm{d}r\mathrm{d}s\\
				&= -\frac{4m^{2} l^{2}\alpha}{2\pi}     \int_{0}^{\infty} \int_{0}^{\infty} e^{-(R+S)m^{2} l^{2}}  \frac{ \left[2\tanh(R)\tanh(S)- \tanh^{2}(R)\right] }{\left[\tanh(R)+\tanh(S)\right]}  \frac{\ln V}{V-1}  \mathrm{d}R \mathrm{d}S
			\end{aligned}  
			\label{kappa}
		\end{equation}
		
		\begin{figure}
			\centering
			\includegraphics[width=0.7\linewidth]{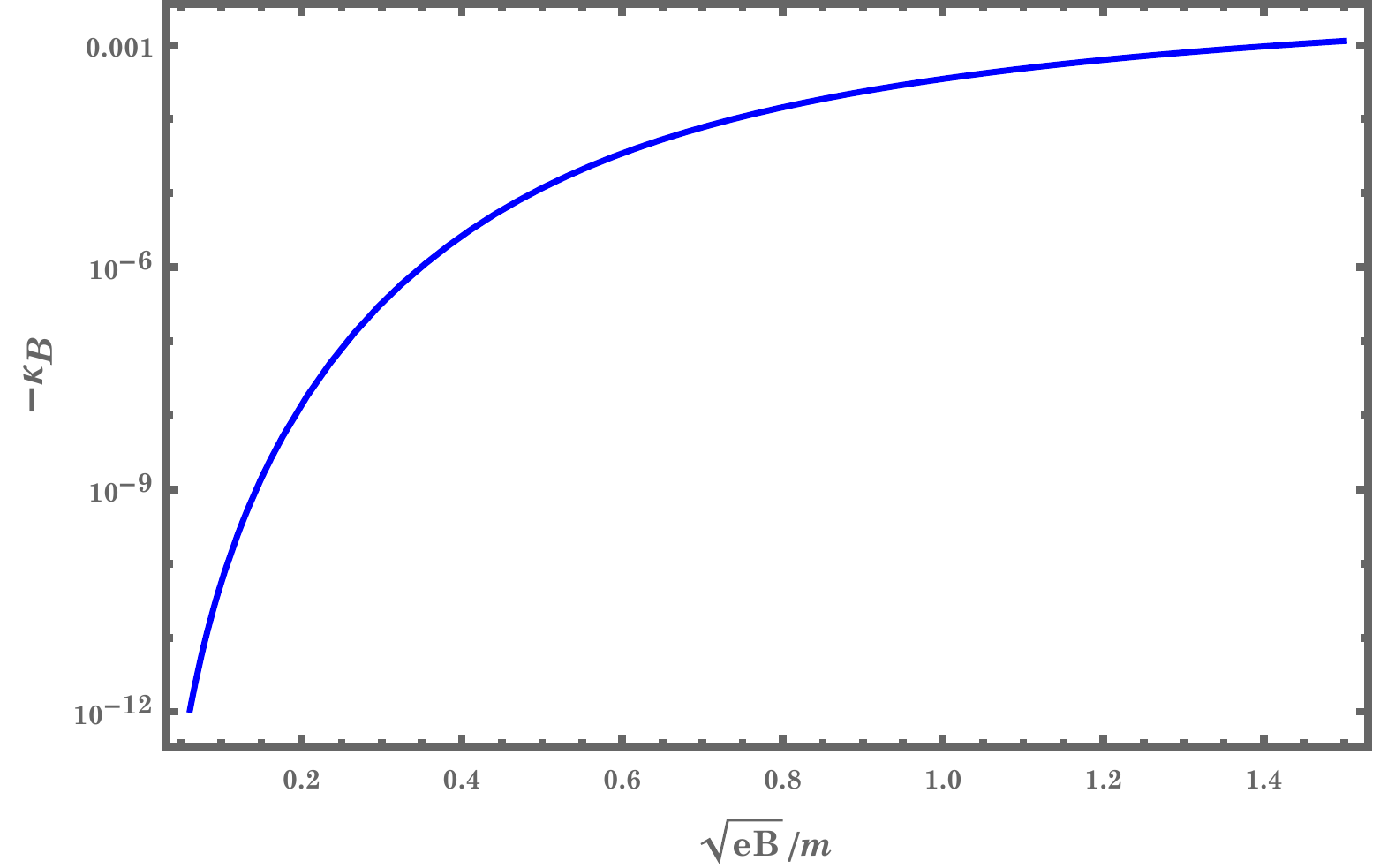}
			\caption{The magnetic correction to the AMM of electron/muon as a function of the dimensionless quantity $\sqrt{eB}/m$, with $m=m_e$ and $m_{\mu}$.}
			\label{MagAMM}
		\end{figure}
		
		\begin{comment}
			\subfigure[]{
				\label{AMMcoElectron}
				\includegraphics[scale=0.5]{electronB.pdf}}
			\subfigure[]{
				\label{AMMcoMuon}
				\includegraphics[scale=0.5]{muonB.pdf}}
		\end{comment}

		with $v=\frac{eBr+eBs}{\tanh(eBr+eBs)} ,V=\frac{R+S}{\tanh(R+S)}, l=1/\sqrt{|eB|}$, and variable transformations $R=eBr,S=eBs$ done.  So we see the final expression of $\kappa_{B}$ manifests the magnetic correction is negative pushing us to check where the minus sign originates. Let's recall the Eq.(\ref{integralI}), there are two electron propagator, one photon propagator, two vertex respectively contributing $(-i)^{2}$, $-i \times (-1)$, $(-i)^{2}$ after Wick rotation with an extra $i$ in differential $\mathrm{d}k$, which results in a minus sign $(-i)^{2}\times i \times (-i)^{2} \times i= -1$.
		Here, we would like to point out that the result in Eq. (\ref{kappa}) cannot continuously goes back to the Schwinger result $a_e=\frac{\alpha}{2\pi}$ when $eB=0$, one can regard this magnetic correction is an extra contribution from magnetic field, and the leading order correction to AMM of electron is: 
		\begin{equation}
			\begin{aligned}
				a_e &=\frac{\alpha}{2\pi}+\kappa_B,\\
				& =\frac{\alpha}{2\pi}\left(1-4m^{2} l^{2}    \int_{0}^{\infty} \int_{0}^{\infty} e^{-(R+S)m^{2} l^{2}}  \frac{ \left[2\tanh(R)\tanh(S)- \tanh^{2}(R)\right] }{\left[\tanh(R)+\tanh(S)\right]}  \frac{\ln V}{V-1}  \mathrm{d}R \mathrm{d}S\right),
			\end{aligned}  
		\end{equation}Analytically,  $m^{2}l^{2}$ is the only dimensionless combination with $eB, m$, which appears twice in the expression: 1) $4m^{2}l^{2}$ as an coefficient in front of the whole expression confirms that the magnetic correction to AMM $\kappa_B \propto m^{2}$ indicates that the magnetic correction is a nonperturbative contribution similar to the AMM induced by the chiral symmetry breaking \cite{Chang:2010hb,Ferrer:2008dy,Ferrer:2013noa}, this also naturally explains why the magnetic correction results cannot smoothly go back to $\frac{\alpha}{2\pi}$ when $eB=0$. The nonperturbative nature of chiral symmetry breaking induced by a constant magnetic field has also been discussed in 1990s \cite{Gusynin:1994re,Gusynin:1994va,Gusynin:1994xp,Gusynin:1995gt}.~2)$m^{2}l^{2}$ shows up in the exponential factor which strongly depresses the integral so that the correction $\kappa_B$  will only become considerable when $\sqrt{eB} \sim m$. 
		
		For the magnetic field magnitude $B \sim  1 ~\mathrm{T} \leftrightarrow eB \sim 2\times 10^{-10} \mathrm{MeV}^{2}$ used in laboratory to measure the electron $g-2$, $|\kappa_B| < 10^{-33}$, much smaller than the measurement $a_{e}=0.001 159 652 180 73 (28)$ \cite{Muong-2:2021ojo,Hanneke:2008tm}. For the muon, whose mass is much heavier than the electron $(m_{\mu}/m_{e})^{2}=43000$,
		under the same magnetic field magnitude $B \sim  1 ~\mathrm{T} \leftrightarrow eB \sim 2\times 10^{-10} \mathrm{MeV}^{2}$, we find that the corresponding magnetic correction to the AMM is $|\kappa_B|< 10^{-42}$, which is also much smaller than the experimental measurement of muon $a_{\mu}=0.00116592040(54)$ \cite{Muong-2:2021ojo}. Thus one can safely neglect the magnetic correction to AMM of electron/muon in the vacuum.
		
		The magnetic correction $\kappa_B$ will become considerable when $\sqrt{eB} \sim m$, which can be seen explicitly from the numerical result shown in Fig.(\ref{MagAMM}) on the magnetic correction $\kappa_B$ in Eq.(\ref{kappa}) to the electron and muon as a function of $\sqrt{eB}/m$ with $m=m_e$ and $m=m_{\mu}$.  To see the negative magnetic correction $\kappa_{B}$ how affect  physics , we consider the electron or a fermion with AMM $\kappa +\kappa_B$ travels in the magnetic field, one can derive its dispersion as below:
		\begin{equation}
			E_{l,s}^{2}=\left\{
			\begin{array}{ll}
				p_{z}^{2}+\left[\sqrt{m^{2}+\left(2l+1+s\right)|eB|}-s(\kappa+\kappa_B) eB\right]^{2}, & l \geq 1 .\\
				&   \\\label{key}
				p_{z}^{2}+[m-(\kappa+\kappa_B) eB]^{2}, &  l=0 .
			\end{array}
			\right.
		\end{equation}
		where $l,s=\pm 1$ are quantum numbers of orbital and spin angular momentum respectively, $eB>0$ for convenience. As is seen, the normal AMM $\kappa$ causes the  splitting of degenerate energy levels known as Zeeman effect in the spectrum, which is  used in accuracy measurement of the AMM \cite{Aoyama:2007dv,Aoyama:2012wj}. However, the energy level splitting will be suppressed with a negative magnetic correction taken into account $\kappa \rightarrow \kappa+\kappa_B$, which will lead to a slight diamagnetism. $\kappa_B$ increases the energy of system expressing a trivial magnetic catalysis effect just as usual $\cite{Miransky:2015ava}$. We will explicitly show these properties in a separate paper \cite{LinfanXukun}.
		
		As addressed in the introduction that the AMM of quark plays a vital role for QCD matter under strong magnetic field when the magnitude of the magnetic field becomes comparable with the quark mass square, such as the early evolution of cosmological phase transition \cite{Vachaspati:1991nm}, non-central heavy ion collisions \cite{Deng:2012pc}, neutron-star merger \cite{Kiuchi:2015sga}. The exact value of AMM of quarks under magnetic field keep unknown because of the nonperturbative nature. In most literature, the AMM of quarks under magnetic field was taken as a positive constant. We can extend the magnetic field dependent AMM of electron to the case of quarks and investigate its effect on the magnetized QCD matter, we will report the results in the coming work \cite{LinfanXukun}.
		
		\vspace*{2mm}
		\section*{Acknowledgements}
		We thank Xinyang Wang for helpful discussions. This work is supported by the NSFC
		under Grant Nos. 11725523 and 11735007, Chinese Academy of Sciences under Grant No.
		XDPB09, the start-up funding from University of Chinese Academy of Sciences(UCAS), and
		the Fundamental Research Funds for the Central Universities.

		\section*{Appendix}
		
		\subsection*{A. Calculation details in section 2.1}
		\appendix
		\setcounter{equation}{0}
		\renewcommand\theequation{A.\arabic{equation}}

		In order to to manage the Schwinger phase $\Phi(x, y) \Phi(y, z)$, we have
		\begin{equation}
			\int_{x}^{y} \xi^{i} \mathrm{d}\xi^{j}=\frac{1}{2}(x+y)^{i}(y-x)^{j} \quad \Longrightarrow \quad \int_{x}^{y} A_{\mu} \mathrm{d}\xi^{\mu}=\frac{B}{2}\epsilon_{ij} x^{j} y^{i} \quad (i,j=1,2),
		\end{equation}
		and the only non-zero terms of $F_{\mu\nu}$ are $ F_{12}=-F_{21}=-B $, which gives
		\begin{equation}
			\int_{x}^{y} \frac{1}{2} F_{\mu \nu}\left(\xi-x\right)^{\nu} \mathrm{d}\xi^{\mu}=-\frac{B}{2}\epsilon_{ij} \frac{(y-x)^{i}(y-x)^{j}}{2}=0.
		\end{equation}
		Then 
		\begin{equation}
			\begin{aligned}
				P(t,k,k^{'})&=\int \mathrm{d}^{4} x \mathrm{d}^{4} y \mathrm{d}^{4} z \, \Phi(x, y) \Phi(y, z) e^{-i t \cdot(z-x)} e^{-i k \cdot(x-y)} e^{-i k^{'} \cdot(y-z)} e^{i p \cdot x} e^{i q \cdot y} e^{-i p^{'}\cdot z}\\
				&=\int \mathrm{d}^{4} h \mathrm{d}^{4} y \mathrm{d}^{4} z \, e^{i\cdot \frac{eB}{2} \epsilon_{ij} y^{i} h^{j}} e^{i h\cdot(p+t-k)} e^{iy\cdot(q+k-k^{'})} e^{iz\cdot(p-p^{'}-k+k^{'})}\\
				&= (2\pi)^{4}\delta^{4}(p-k+k^{'}-p^{'}) \int \mathrm{d}^{4} h \mathrm{d}^{4} y \, e^{i\cdot \frac{eB}{2} \epsilon_{ij} y^{i} h^{j}} e^{i h\cdot(p+t-k)} e^{iy\cdot(q+k-k^{'})}\\
				&=(2\pi)^{4}\delta^{4}(p-k+k^{'}-p^{'}) \int \mathrm{d}^{2} h_{\|} \mathrm{d}^{2} y_{\|} \,  e^{i h_{\|}\cdot(p+t-k)_{\|}} e^{iy_{\|}\cdot(q+k-k^{'})_{\|}} \\
				&\times  \int \mathrm{d}^{2} h_{\perp} \mathrm{d}^{2} y_{\perp} \,  e^{i h_{\perp}\cdot(p+t-k)_{\perp}} e^{iy_{\perp}\cdot(q+k-k^{'})_{\perp}} \times e^{i\cdot \frac{eB}{2} \epsilon_{ij} y^{i} h^{j}}\\
				& =(2\pi)^{4}\delta^{4}(p-k+k^{'}-p^{'})(2\pi)^{2}\delta^{2}(p_{\|}+t_{\|}-k_{\|})(2\pi)^{2}\delta^{2}(q_{\|}+k_{\|}-k^{'}_{\|})\\
				& \times \int \mathrm{d}^{2} h_{\perp} \mathrm{d}^{2} y_{\perp} \,  e^{i h_{\perp}\cdot(p+t-k)_{\perp}} e^{iy_{\perp}\cdot(q+k-k^{'})_{\perp}} \times e^{i\cdot \frac{eB}{2} \epsilon_{ij} y^{i} h^{j}}\\
				&=(2\pi)^{4}\delta^{4}(p-k+k^{'}-p^{'})(2\pi)^{2}\delta^{2}(p_{\|}+t_{\|}-k_{\|})(2\pi)^{2}\delta^{2}(q_{\|}+k_{\|}-k^{'}_{\|})\\
				&\times \int \mathrm{d}^{2} h_{\perp} \, e^{i h_{\perp}\cdot(p+t-k)_{\perp}}  \int \mathrm{d} y^{1} d y^{2} \,  e^{iy^{1} \left((q+k-k^{'})_{1}-\frac{eB}{2}h_{2}\right)} e^{iy^{2} \left((q+k-k^{'})_{2}-\frac{eB}{2}h_{1}\right)}\\
				&=(2\pi)^{4}\delta^{4}(p-k+k^{'}-p^{'})(2\pi)^{2}\delta^{2}(p_{\|}+t_{\|}-k_{\|})(2\pi)^{2}\delta^{2}(q_{\|}+k_{\|}-k^{'}_{\|})\\
				& \times \int \mathrm{d}^{2} h_{\perp} \, e^{i h_{\perp}\cdot(p+t-k)_{\perp}} (2\pi)^{2} \delta\left((q+k-k^{'})_{1}-\frac{eB}{2}h_{2}\right) \delta \left((q+k-k^{'})_{2}-\frac{eB}{2}h_{1}\right)\\
				&=(2\pi)^{4}\delta^{4}(p-k+k^{'}-p^{'})(2\pi)^{2}\delta^{2}(p_{\|}+t_{\|}-k_{\|})(2\pi)^{2}\delta^{2}(q_{\|}+k_{\|}-k^{'}_{\|})\\
				& \times (2\pi)^{2} 4l^{4} e^{-i\cdot2l^{2}(p+t-k)^{1}(q+k-k^{'})_{2}} e^{i\cdot 2l^{2}(p+t-k)^{2}(q+k-k^{'})_{1}},
			\end{aligned}
		\end{equation}
		where the new conventions are defined: $p_{\|}\cdot k_{\|}\equiv p^{0}k^{0}-p^{3}k^{3},p_{\perp}\cdot k_{\perp}\equiv -p^{1}k^{1}-p^{2}k^{2}$, so $p \cdot k =p_{\|}\cdot k_{\|}+ p_{\perp}\cdot k_{\perp}$. we can integrate $t_{||},k^{'}_{||}$:
		\begin{equation}
			\begin{aligned}
				\mathcal{M}^{\mu} &=\int \frac{\mathrm{d}^{4} t}{(2 \pi)^{4}} \frac{\mathrm{d}^{4} k^{'}}{(2 \pi)^{4}} \frac{\mathrm{d}^{4} k}{(2 \pi)^{4}} P(t,k,k^{'})\times\mathcal{A}^{\mu}(t,k,k{'})\\
				&=\int (\mathrm{d}t \mathrm{d}k^{'} \mathrm{d}k) (2\pi)^{4}\delta^{4}(p-k+k^{'}-p^{'})(2\pi)^{2}\delta^{2}(p_{\|}+t_{\|}-k_{\|})(2\pi)^{2}\delta^{2}(q_{\|}+k_{\|}-k^{'}_{\|})\\
				& \times (2\pi)^{2} 4l^{4} e^{-i\cdot 2l^{2}(p+t-k)^{1}(q+k-k^{'})_{2}} e^{i\cdot 2l^{2}(p+t-k)^{2}(q+k-k^{'})_{1}} \times\mathcal{A}^{\mu}(t,k,k{'})\\
				& = \int (\mathrm{d}t_{\perp} \mathrm{d}k^{'}_{\perp}\mathrm{d}k) (2\pi)^{2}\delta^{2}(p^{'}_{\|}-p_{\|}=q_{\|}) (2\pi)^{2}\delta^{2}(p_{\perp}-k_{\perp}=p^{'}_{\perp}-k^{'}_{\perp}) \\
				& \times (2\pi)^{2} 4l^{4} e^{-i\cdot 2l^{2}(p+t-k)^{1}(q+k-k^{'})_{2}} e^{i\cdot 2l^{2}(p+t-k)^{2}(q+k-k^{'})_{1}} \times\mathcal{A}^{\mu}(t,k,k{'}).
			\end{aligned}
		\end{equation}

		\subsection*{B. Calculation details in section 2.2}
		\appendix
		\setcounter{equation}{0}
		\renewcommand\theequation{B.\arabic{equation}}
		
		After Wick rotation $k^{0} \rightarrow ik^{0},r \rightarrow -ir,s \rightarrow -is$, we obtain
		\begin{equation}
			\begin{aligned}
				\mathcal{I} &=(2\pi)^{2} 4l^{4} \times (-i)^{2} \int_{0}^{\infty} \mathrm{d}r \int_{0}^{\infty} \mathrm{d}s  \,\times (i) \int (\mathrm{d}t_{\perp} \mathrm{d}k^{'}_{\perp} \mathrm{d}k_{E}) \frac{-i e^2}{k^{2}_{\|E}+t^{2}_{\perp}}  \\
				& \times e^{-i\cdot 2l^{2}(t-k)^{1}(k-k^{'})_{2}} e^{i\cdot 2l^{2}(t-k)^{2}(k-k^{'})_{1}}  e^{-r\left(k_{\|E}^{2}+m^{2}\right)-k_{\perp}^{'2} l^{2} \tanh \left(e Br\right)}  e^{-s\left(k_{\|E}^{2}+m^{2}\right)-k_{\perp}^{2} l^{2} \tanh \left(e Bs\right)},
			\end{aligned}
		\end{equation}
		the subscript $ _{E}$ will be omitted from now for convenience. Firstly, we do the integral related to $t_{\perp}$:
		\begin{eqnarray}
			\mathcal{I} &= &\left[-e^{2} (2\pi)^{2} 4l^{4} \right] \int_{0}^{\infty} \mathrm{d}r \int_{0}^{\infty} \mathrm{d}s \int(\mathrm{d}k^{'}_{\perp}\mathrm{d}k) \nonumber \\
			& &	e^{-r\left(k_{\|}^{2}+m^{2}\right)-k_{\perp}^{'2} l^{2} \tanh \left(e Br\right)}  e^{-s\left(k_{\|}^{2}+m^{2}\right)-k_{\perp}^{2} l^{2} \tanh \left(e Bs\right)}  \times \mathcal{T}_{\perp},
		\end{eqnarray}
		where
		\begin{equation}
			\begin{aligned}
				\mathcal{T}_{\perp}&=\int (\mathrm{d}t_{\perp}) \, \frac{1}{k^{2}_{\|}+t^{2}_{\perp}}  e^{i\cdot 2l^{2}(t-k)^{1}(k-k^{'})^{2}} e^{-i\cdot 2l^{2}(t-k)^{2}(k-k^{'})^{1}} \\
				& = e^{-i\cdot 2l^{2} k^{1}(k-k^{'})^{2}} e^{i\cdot 2l^{2} k^{2}(k-k^{'})^{1}} \int \frac{\mathrm{d}t^{1}\mathrm{d}t^{2}}{(2 \pi)^2} \, e^{-iat^{1}} e^{-ibt^{2}} \frac{1}{k^{2}_{\|}+t_{1}^{2}+t_{2}^{2}}\\
				&=  e^{-i\cdot 2l^{2} k^{1}(k-k^{'})^{2}} e^{i\cdot 2l^{2} k^{2}(k-k^{'})^{1}} \int^{+\infty}_{-\infty}\frac{\mathrm{d}t^{2}}{2 \pi} \, e^{-ibt^{2}} \frac{e^{-\sqrt{k^{2}_{\|}+t_{2}^{2}} |a|}}{2\sqrt{k^{2}_{\|}+t_{2}^{2}}}\\
				&=  e^{-i\cdot 2l^{2} k^{1}(k-k^{'})^{2}} e^{i\cdot 2l^{2} k^{2}(k-k^{'})^{1}} \frac{1}{2\pi }\int^{+\infty}_{0} \mathrm{d}t^{2} \,\frac{e^{-\sqrt{k^{2}_{\|}+t_{2}^{2}} |a|}}{\sqrt{k^{2}_{\|}+t_{2}^{2}}} \cos(bt^{2})\\
				&= e^{-i\cdot 2l^{2} k^{1}(k-k^{'})^{2}} e^{i\cdot 2l^{2} k^{2}(k-k^{'})^{1}} \frac{1}{2\pi } K_{0}(|k_{\|}|\sqrt{a^2+b^2}).
			\end{aligned}
		\end{equation}
		where $a=-2 l^{2} (k-k^{'})^{2},b=2 l^{2}(k-k^{'})^{1}$ and $K_{\nu}(z)$ is the Bessel function of imaginary argument, two formulae have been used \cite{Gradshteyn:2014} in the calculation:
		\begin{equation}
			\frac{1}{2\pi} \int_{-\infty}^{+\infty} \frac{ e^{-i\xi x}}{c^{2}+x^{2}} \mathrm{d}x=\frac{e^{-c|\xi|}}{2c}  \quad  [c>0,\xi \in \mathbb{R}],   
		\end{equation}
		\begin{equation}
			\int_{0}^{\infty} \frac{e^{-\beta  \sqrt{\gamma^{2}+x^{2}}}}{\sqrt{\gamma^{2}+x^{2}}} \cos\! b x \, \mathrm{d} x=K_{0}\left(\gamma \sqrt{\beta^{2}+b^{2}}\right) \quad[\operatorname{Re} \beta >0, \text { Re } \gamma>0,  b>0].
		\end{equation}
		Then, we obtain
		\begin{equation}
			\begin{aligned}
				\mathcal{I} &= \int_{0}^{\infty} \mathrm{d}r \int_{0}^{\infty} \mathrm{d}s \int(\mathrm{d}k^{'}_{\perp}\mathrm{d}k) 
				e^{-r\left(k_{\|}^{2}+m^{2}\right)-k_{\perp}^{'2} l^{2} \tanh \left(e Br\right)}  e^{-s\left(k_{\|}^{2}+m^{2}\right)-k_{\perp}^{2} l^{2} \tanh \left(e Bs\right)}\\
				& \times \left[-e^{2} (2\pi)^{2} 4l^{4} \right]  e^{-i\cdot 2l^{2} k^{1}(k-k^{'})^{2}} e^{i\cdot 2l^{2} k^{2}(k-k^{'})^{1}} \frac{1}{2\pi } K_{0}(|k_{\|}|\sqrt{a^2+b^2}).
			\end{aligned}
		\end{equation}
		Making a variable transformation  $\tilde{k}_{\perp}=k^{'}_{\perp}-k_{\perp},d\tilde{k}_{\perp}=dk^{'}_{\perp}$, it is easy to see $\sqrt{a^{2}+b^{2}}=2l^{2}\sqrt{(k-k^{'})^{2}_{\perp}}=2 l^{2} |\tilde{k}_{\perp}|$, which simplifies the expression:
		\begin{equation}
			\begin{aligned}
				\mathcal{I} &= \left[-e^{2} (2\pi) 4l^{4} \right]\int_{0}^{\infty} \mathrm{d}r \int_{0}^{\infty} \mathrm{d}s \int(\mathrm{d}\tilde{k}_{\perp}\mathrm{d}k) 
				e^{-i\cdot 2l^{2} k^{1}(k-k^{'})^{2}} e^{i\cdot 2l^{2} k^{2}(k-k^{'})^{1}} \\
				& \times e^{-r\left(k_{\|}^{2}+m^{2}\right)-(\tilde{k}_{\perp}+k_{\perp})^{2} l^{2} \tanh \left(e Br\right)}  e^{-s\left(k_{\|}^{2}+m^{2}\right)-k_{\perp}^{2} l^{2} \tanh \left(e Bs\right)} K_{0}(2 l^{2} |k_{\|}| |\tilde{k}_{\perp}|) \\
				& =\left[-e^{2} (2\pi) 4l^{4} \right]\int_{0}^{\infty} \mathrm{d}r \int_{0}^{\infty} \mathrm{d}s \int(\mathrm{d}\tilde{k}_{\perp} \mathrm{d} k_{\|})  e^{-(r+s)\left(k_{\|}^{2}+m^{2}\right)-\tilde{k}_{\perp}^{2} l^{2} \tanh \left(e Br\right)} K_{0}(2 l^{2} |k_{\|}| |\tilde{k}_{\perp}|) \times \mathcal{K}_{\perp},
			\end{aligned}
		\end{equation}
		where $\mathcal{K}_{\perp}$ is the part related to $k_{\perp}$:
		\begin{equation}
			\begin{aligned}
				\mathcal{K}_{\perp}&=\int (\mathrm{d} k_{\perp})  e^{i\cdot 2l^{2} k_{1}\tilde{k}_{2}} e^{-i\cdot 2l^{2} k_{2}\tilde{k}_{1}}  e^{-k_{\perp}^{2} l^{2} \tanh \left(e Br\right)-2 l^{2} \tilde{k}_{\perp}\cdot k_{\perp} \tanh(eBr)} e^{-k_{\perp}^{2} l^{2} \tanh \left(e Bs\right)} \\
				& =\int \frac{\mathrm{d}k_{1}}{2\pi} e^{i \cdot 2l^{2} k_{1} \tilde{k}_{2}} e^{-k_{1}^{2}l^{2}\tanh(eBr)-k_{1}^{2}l^{2}\tanh(eBs)-2k_{1}\tilde{k}_{1}\tanh(eBr)}    \\
				& \times \int \frac{\mathrm{d}k_{2}}{2\pi}  e^{-i \cdot 2l^{2} k_{2} \tilde{k}_{1}} e^{-k_{2}^{2}l^{2}\tanh(eBr)-k_{2}^{2}l^{2}\tanh(eBs)-2k_{2}\tilde{k}_{2}\tanh(eBr)} \\
				& = \frac{1}{(2\pi)^{2}}   \frac{\pi}{l^{2}\left[\tanh(eBr)+\tanh(eBs)\right]} e^{\frac{4l^{4}\left[\tilde{k}_{1}\tanh(eBr)-i\tilde{k}_{2}\right]^{2}}  {4l^{2}\left[\tanh(eBr)+\tanh(eBs)\right]} }  e^{  \frac{4l^{4}\left[\tilde{k}_{2}\tanh(eBr)-i\tilde{k}_{1}\right]^{2}}  {4l^{2}\left[\tanh(eBr)+\tanh(eBs)\right]} } \\
				& =\frac{1}{4\pi l^{2}}   \frac{1}{\left[\tanh(eBr)+\tanh(eBs)\right]}  \exp{  \left\{    \frac{\tilde{k}_{\perp}^{2}  l^{2} \left[\tanh^{2}(eBr)-1\right]}  {\tanh(eBr)+\tanh(eBs)}    \right\}  },
			\end{aligned}
		\end{equation}
		an integral formula has been used \cite{Gradshteyn:2014}:
		\begin{equation}
			\int_{-\infty}^{\infty} \exp \left(-p^{2} x^{2} \pm q x\right) \mathrm{d} x=\exp \left(\frac{q^{2}}{4 p^{2}}\right) \frac{\sqrt{\pi}}{p} \quad\left[\operatorname{Re} p^{2}>0\right].
		\end{equation}
		We have 
		\begin{equation}
			\begin{aligned}
				\mathcal{I} &= \int_{0}^{\infty} \mathrm{d}r \int_{0}^{\infty} \mathrm{d}s \int(\mathrm{d}\tilde{k}_{\perp} \mathrm{d}k_{\|}) \left[-e^{2} (2\pi) 4l^{4} \right] e^{-(r+s)\left(k_{\|}^{2}+m^{2}\right)-\tilde{k}_{\perp}^{2} l^{2} \tanh \left(e Br\right)}\\
				& \times \frac{1}{4\pi l^{2} }   \frac{K_{0}(2 l^{2} |k_{\|}| |\tilde{k}_{\perp}|) }{\left[\tanh(eBr)+\tanh(eBs)\right]}  \exp{  \left\{   \frac{\tilde{k}_{\perp}^{2}  l^{2} \left[\tanh^{2}(eBr)-1\right]}  {\tanh(eBr)+\tanh(eBs)}   \right\}  } \\
				& = \int_{0}^{\infty} \mathrm{d}r \int_{0}^{\infty} \mathrm{d}s \int(\mathrm{d}\tilde{k}_{\perp} \mathrm{d} k_{\|}) \left[-e^{2} 2l^{2} \right] 
				e^{-(r+s)\left(k_{\|}^{2}+m^{2}\right)-\frac{\tilde{k}_{\perp}^{2}  l^{2} }  {\tanh(eBr+eBs)} }\\
				& \times \frac{1 }{\left[\tanh(eBr)+\tanh(eBs)\right]} \times \frac{1}{2} \int_{0}^{\infty} \mathrm{d}u \frac{e^{-u}}{u} e^{-\frac{l^{4}k^{2}_{\|}  \tilde{k}^{2}_{\perp}}{u}} \\
				& =  \int_{0}^{\infty} \mathrm{d}r \int_{0}^{\infty} \mathrm{d}s \frac{-e^{2} l^{2} }{\left[\tanh(eBr)+\tanh(eBs)\right]}   \int(\mathrm{d} k_{\|}) e^{-(r+s)\left(k_{\|}^{2}+m^{2}\right)} \times \mathcal{U},
			\end{aligned}  
		\end{equation}
		where \begin{equation}
			\mathcal{U} = \int_{0}^{\infty} \mathrm{d}u \frac{e^{-u}}{u} \int(\mathrm{d}\tilde{k}_{\perp})  e^{-\frac{\tilde{k}_{\perp}^{2}  l^{2} }  {\tanh(eBr+eBs)}  -\frac{l^{4}k^{2}_{\|}  \tilde{k}^{2}_{\perp}}{u} },
		\end{equation}
		the formulae we used here are \cite{Gradshteyn:2014}: 
		\begin{equation}
			K_{0}(z)=\frac{1}{2}\int_{0}^{\infty} \frac{e^{-u-z^{2} / 4 u} \mathrm{d} u}{u} \quad\left[|\arg z|<\frac{\pi}{4}, \quad \operatorname{Re} z^{2}>0\right],
		\end{equation}
		\begin{equation}
			\tanh (x+y)=\frac{\tanh x+\tanh y}{1+\tanh x \tanh y}.
		\end{equation}
		Define two new notations $\alpha, \beta $:
		\begin{equation}
			\alpha+\frac{\beta}{u}=\frac{l^{2} }  {\tanh(eBr+eBs)}  +\frac{l^{4}k^{2}_{\|} }{u},
		\end{equation}
		so that we can integrate the $\tilde{k}_{\perp}$ :
		\begin{equation}
			\begin{aligned}
				\mathcal{U} &= \int_{0}^{\infty} \mathrm{d}u \frac{e^{-u}}{u} \int(\mathrm{d}\tilde{k}_{\perp})  e^{-\frac{\tilde{k}_{\perp}^{2}  l^{2} }  {\tanh(eBr+eBs)}  -\frac{l^{4}k^{2}_{\|}  \tilde{k}^{2}_{\perp}}{u} }= \int_{0}^{\infty} \mathrm{d}u \frac{e^{-u}}{u} \int(\mathrm{d}\tilde{k}_{\perp}) e^{-(\alpha+\frac{\beta}{u})\tilde{k}_{\perp}^{2}}\\
				& = \int_{0}^{\infty} \mathrm{d}u \frac{e^{-u}}{u} \int^{\infty}_{0} \frac{\mathrm{d}|\tilde{k}_{\perp}|}{2\pi}  |\tilde{k}_{\perp}| e^{-(\alpha+\frac{\beta}{u}) |\tilde{k}|_{\perp}^{2}}  = \frac{1}{4\pi} \int_{0}^{\infty} \mathrm{d}u \frac{e^{-u}}{\alpha u+\beta} \\
				& =\frac{1}{4\pi} \int_{0}^{\infty} \mathrm{d}u  \frac{e^{-u}}{\alpha u+l^{4}k^{2}_{\|}},
			\end{aligned}
		\end{equation}
		which simplifies $\mathcal{I}$ again:
		\begin{equation}
			\begin{aligned}
				\mathcal{I} &=  \int_{0}^{\infty} \int_{0}^{\infty}\frac{\left[-e^{2} l^{2} \right] \mathrm{d}r \mathrm{d}s }{\left[\tanh(eBr)+\tanh(eBs)\right]}   \int(\mathrm{d} k_{\|}) e^{-(r+s)\left(k_{\|}^{2}+m^{2}\right)}  \times  \frac{1}{4\pi} \int_{0}^{\infty} du \frac{e^{-u}}{\alpha u+l^{4}k^{2}_{\|}} \\
				& = \frac{-e^{2} l^{2}}{4\pi}  \int_{0}^{\infty} \int_{0}^{\infty}\frac{  e^{-(r+s)m^{2}}  \mathrm{d}r \mathrm{d}s }{\left[\tanh(eBr)+\tanh(eBs)\right]}  
				\int(\mathrm{d} k_{\|}) e^{-(r+s)k_{\|}^{2}} \int_{0}^{\infty} \mathrm{d}u \frac{e^{-u}}{\alpha u+l^{4}k^{2}_{\|}} \\
				& =\frac{-e^{2} l^{2}}{4\pi}  \int_{0}^{\infty} \int_{0}^{\infty}\frac{  e^{-(r+s)m^{2}}  \mathrm{d}r \mathrm{d}s }{\left[\tanh(eBr)+\tanh(eBs)\right]} \times \mathcal{K}_{\|}.
			\end{aligned}  
		\end{equation}
		The integral of $k_{||}$ becomes:
		\begin{equation}
			\begin{aligned}
				\mathcal{K}_{\|} &= \int(\mathrm{d} k_{\|}) e^{-(r+s)k_{\|}^{2}} \int_{0}^{\infty}\mathrm{d}u \frac{e^{-u}}{\alpha u+l^{4}k^{2}_{\|}}  \\
				&= \int_{0}^{\infty} \mathrm{d}u e^{-u} \int_{0}^{\infty} \frac{\mathrm{d} | k_{\|} |}{2\pi l^{4}} 
				\frac{| k_{\|}   e^{-(r+s) |k_{\|}|^{2}}  }{\alpha u /l^{4} + |k_{\|}|^{2}} \\
				& =\frac{1}{4\pi l^{4}}  \int_{0}^{\infty} \mathrm{d}u  \int_{0}^{\infty} \mathrm{d}w \frac{e^{-w-u}}{w+v u} \\
				& = \frac{1}{4\pi l^{4}} \frac{\ln v}{v-1},
			\end{aligned}
		\end{equation}
		where a variable transformation is applied $w= |k_{||}|^{2}$, and $v=(r+s)\alpha/l^{4}=eB(r+s)/\tanh(eBr+eBs) \ge 1 $. Here a necessary  integral is used:
		\begin{eqnarray}
			\int_{0}^{\infty} \frac{x e^{-ax^{2}}}{b+x^{2}} \mathrm{d}x &=& \frac{1}{2}  \int_{0}^{\infty} \frac{e^{-w} \mathrm{d}w}{w+ab} \nonumber \\
			&=& \frac{1}{2} e^{ab} \Gamma(0,ab) \quad [ a,b>0 ],
		\end{eqnarray}
		$\Gamma(x, z)=\int_{z}^{\infty} t^{x-1} e^{-t} \mathrm{d} t$ is the incomplete Gamma function. Through tedious computation, we finally receive:
		\begin{equation}
			\begin{aligned}
				\mathcal{I} &= \frac{-e^{2} l^{2}}{4\pi}  \int_{0}^{\infty} \int_{0}^{\infty}\frac{  e^{-(r+s)m^{2}}  \mathrm{d}r \mathrm{d}s }{\left[\tanh(eBr)+\tanh(eBs)\right]}  
				\times \frac{1}{4\pi l^{4}} \frac{\ln v}{v-1} \\
				& = -\frac{\alpha_{\mathrm{QED}}}{4\pi l^{2}}\int_{0}^{\infty} \int_{0}^{\infty}\frac{  e^{-(r+s)m^{2}}  \mathrm{d}r \mathrm{d}s }{\left[\tanh(eBr)+\tanh(eBs)\right]}  \frac{\ln v}{v-1}
			\end{aligned}  
		\end{equation}
		with $v=\frac{eBr+eBs}{\tanh(eBr+eBs)}$ .

		\subsection*{C. Calculation details in section 2.3}
		\appendix
		\setcounter{equation}{0}
		\renewcommand\theequation{C.\arabic{equation}}
		\par When $\mu=1$, we can simplify the expression into a more friendly form by Dirac's  algebra: 
		\begin{equation}
			\begin{aligned}
				\Theta^{\mu}(r,s)_{\mathrm{AMM}}&=4 m \tilde{q}_{\alpha}\left[g^{\alpha 1}+g^{\alpha 2}\tan(eBs) -g^{\alpha 2}\tan(eBr) +g^{\alpha 1}\tan(eBr)\tan(eBs)\right]\\
				&=4 m \tilde{q}_{\alpha}\left[i \sigma^{\alpha 1}+i \sigma^{\alpha 2}\tan(eBs)
				-i \sigma^{\alpha 2}\tan(eBr)+i \sigma^{\alpha 1}\tan(eBr)\tan(eBs)\right] \\
				&+ 4 m \tilde{q}_{\alpha}\left[\gamma^{\alpha}\gamma^{1}+\gamma^{\alpha}\gamma^{2}\tan(eBs)
				-\gamma^{\alpha}\gamma^{2}\tan(eBr)+\gamma^{\alpha}\gamma^{1}\tan(eBr)\tan(eBs)\right]\\
				&=\frac{i\sigma^{1\alpha} q_{\alpha}}{2m}\{-8m^{2}\left[1+\tan(eBr)\tan(eBs)\right]\} 
				+\frac{i\sigma^{2\alpha} q_{\alpha}}{2m}\{-8m^{2}\left[\tan(eBs)-\tan(eBr)\right]\}\\
				&+\frac{i\sigma^{1\alpha}\hat{q}_{\alpha}}{2m}\{-8m^{2}\left[1+\tan(eBr)\tan(eBs)\right]\} +\frac{i\sigma^{2\alpha}\hat{q}_{\alpha}}{2m}\{-8m^{2}\left[\tan(eBs)-\tan(eBr)\right]\}\\
				&+ 4 m \slashed{q}\left[\gamma^{1}+\gamma^{2}\tan(eBs)-\gamma^{2}\tan(eBr)+\gamma^{1}\tan(eBr)\tan(eBs)\right]\\
				&+4m\slashed{\hat{q}}\left[\gamma^{1}+\gamma^{2}\tan(eBs)-\gamma^{2}\tan(eBr)+\gamma^{1}\tan(eBr)\tan(eBs)\right] ,
			\end{aligned}
			\label{mu1}
		\end{equation}
		Similarly, when $\mu=2$
		\begin{equation}
			\begin{aligned}
				\Theta^{\mu}(r,s)_{\mathrm{AMM}}&=4 m \tilde{q}_{\alpha}\left[g^{\alpha 2}-g^{\alpha 1}\tan(eBs)+g^{\alpha 1}\tan(eBr)+g^{\alpha 2}\tan(eBr)\tan(eBs)\right]\\
				&=4 m \tilde{q}_{\alpha}\left[i \sigma^{\alpha 2}-i \sigma^{\alpha 1}\tan(eBs)
				+i \sigma^{\alpha 1}\tan(eBr)+i \sigma^{\alpha 2}\tan(eBr)\tan(eBs)\right] \\
				&+ 4 m \tilde{q}_{\alpha}\left[\gamma^{\alpha}\gamma^{2}-\gamma^{\alpha}\gamma^{1}\tan(eBs)+\gamma^{\alpha}\gamma^{1}\tan(eBr)+\gamma^{\alpha}\gamma^{2}\tan(eBr)\tan(eBs)\right]\\
				&=\frac{i\sigma^{2\alpha} q_{\alpha}}{2m}\{-8m^{2}\left[1+\tan(eBr)\tan(eBs)\right]\} +\frac{i\sigma^{1\alpha} q_{\alpha}}{2m}\{-8m^{2}\left[\tan(eBr)-\tan(eBs)\right]\}\\
				&+\frac{i\sigma^{2\alpha}\hat{q}_{\alpha}}{2m}\{-8m^{2}\left[1+\tan(eBr)\tan(eBs)\right]\} +\frac{i\sigma^{1\alpha}\hat{q}_{\alpha}}{2m}\{-8m^{2}\left[\tan(eBr)-\tan(eBs)\right]\}\\
				&+ 4 m \slashed{q}\left[\gamma^{2}-\gamma^{1}\tan(eBs)+\gamma^{1}\tan(eBr)+\gamma^{2}\tan(eBr)\tan(eBs)\right]\\
				&+4m\slashed{\hat{q}}\left[\gamma^{2}-\gamma^{1}\tan(eBs)+\gamma^{1}\tan(eBr)+\gamma^{2}\tan(eBr)\tan(eBs)\right],
			\end{aligned}
			\label{mu2}
		\end{equation}	
		with $\tilde{q}_{\mu}=q_{\mu}+\hat{q}_{\mu},\hat{q}_{\mu}:=\left(0,q_{2}\tan(eBr),-q_{1}\tan(eBr),0\right)$.
		Consider the last equal sign in Eq.(\ref{mu1},\ref{mu2}), $\frac{i\sigma^{1\alpha} q_{\alpha}}{2m},\frac{i\sigma^{2\alpha} q_{\alpha}}{2m}$ terms are what we seek providing AMM with a magnetic correction, $4 m \slashed{q}$ terms actually are irrelevant with AMM by a conversion in virtue of Gordon identity:
		\begin{equation}
			\bar{u}\left(p^{\prime}\right)\slashed{q}\gamma^{\mu}u(p)=\bar{u}\left(p^{\prime}\right) \left[2m \gamma^{\mu} -2p^{\mu}\right]u(p),
		\end{equation}
		the initial and final state of  electron are considered in infinitely far positions where is no magnetic field so Gordon identity holds validity. 
		For the remaining terms, the analysis is a little tedious by searching corresponding non-relativistic limit as done in quantum field theory \cite{Peskin:1995ev}:
		\begin{equation}
			\bar{u}(p^{'})\Theta^{i}u(p)\tilde{A}^{i}=\bar{u}(p^{'})\Theta^{1}u(p)\tilde{A}^{1}+\bar{u}(p^{'})\Theta^{2}u(p)\tilde{A}^{2}.
		\end{equation}
		Insert the non-relativistic expansion and keep the leading order of momentum:
		\begin{equation}
			u(p)=\left(\begin{array}{c}
				\sqrt{p \cdot \sigma} \xi \\
				\sqrt{p \cdot \bar{\sigma} }\xi
			\end{array}\right) \approx \sqrt{m}\left(\begin{array}{c}
				(1-\mathbf{p} \cdot \sigma / 2 m) \xi \\
				(1+\mathbf{p} \cdot \sigma / 2 m) \xi
			\end{array}\right).
		\end{equation}
		We have 
		\begin{equation}
			\bar{u}(p^{'})\frac{i\sigma^{1\alpha}\hat{q}_{\alpha}}{2m} u(p)\tilde{A}^{1}+\bar{u}(p^{'})\frac{i\sigma^{2\alpha}\hat{q}_{\alpha}}{2m} u(p)\tilde{A}^{2} \approx i (\hat{q}_{2}\tilde{A}^{1}-\hat{q}_{1}\tilde{A}^{2})\xi^{\prime\dagger}\sigma^{3}\xi,
		\end{equation}
		\begin{equation}
			\bar{u}(p^{'})\frac{i\sigma^{2\alpha}\hat{q}_{\alpha}}{2m} u(p)\tilde{A}^{1}+\bar{u}(p^{'})\frac{i\sigma^{1\alpha}\hat{q}_{\alpha}}{2m} u(p)\tilde{A}^{2} \approx i (\hat{q}_{2}\tilde{A}^{2}-\hat{q}_{1}\tilde{A}^{1})\xi^{\prime\dagger}\sigma^{3}\xi,
		\end{equation}
		so taking $\hat{q}_{\mu}=\left(0,q_{2}\tan(eBr),-q_{1}\tan(eBr),0\right)$ into account, one can find the above two terms are gauge-dependent corresponding to operators $ \partial_{1}A^{1}+\partial_{2}A^{2}$ and $\partial_{1}A^{2}+\partial_{2}A^{1} $ in coordinate space. So the fifth lines of Eq.(\ref{mu1},\ref{mu2}) make no sense for a observable quantity, whose gauge-dependent will eliminate after integrate with $q$ in a complete calculation of Feynman diagram. Similarly, for the last line of Eq.(\ref{mu1},\ref{mu2}), we analyze in the same way and find that the principle contribution to the AMM is 
		\begin{equation}
			4m(\tan(eBs)-\tan(eBr))(\hat{q}_{2}\tilde{A}^{1}+\hat{q}_{1}\tilde{A}^{1})\gamma^{1}\gamma^{2},
		\end{equation}
		after inserting the spinor expansion:
		\begin{equation}
			\begin{aligned}
				&\bar{u}(p^{'})\left[  4m(\tan(eBs)-\tan(eBr))(\hat{q}_{1}\tilde{A}^{1}+\hat{q}_{2}\tilde{A}^{2})\gamma^{1}\gamma^{2}\right] u(p)\\
				\approx & 8m^{2}\tan(eBr)\left[ \tan(eBr)-\tan(eBs)\right] \left[ -i(q^{2}\tilde{A}^{1}-q^{1}\tilde{A}^{2})\right] \xi^{\prime\dagger}\sigma^{3}\xi.
			\end{aligned}
		\end{equation}
		we get the coefficient of total magnetic correction term of electron as following
		\begin{equation}
			\begin{aligned}
				&8m^{2}\tan(eBr)\left[ \tan(eBr)-\tan(eBs)\right]-8m^{2}\left[1+\tan(eBr)\tan(eBs)\right]\\
				=&\,8m^{2}\left[ \tan^{2}(eBr)-2\tan(eBr)\tan(eBs)-1\right] .
			\end{aligned}
		\end{equation}
		Because of the dimension reduction originating from the magnetic field, the leaving term $-8 m^{2}$ after directly setting $eB=0$ in above makes no sense. So our final magnetic correction of AMM is $8m^{2}\left[ \tan^{2}(eBr)-2\tan(eBr)\tan(eBs)\right]$.
		For the longitudinal directions $\mu=0 , 3$:
		\begin{equation}
			\begin{aligned}
				\Theta^{\mu}(r,s)_{\mathrm{AMM}}&= m\gamma^{\nu}\slashed{\tilde{q}}  \left[ \gamma^{\mu}-\gamma^{\mu}\gamma^{1}\gamma^{2}\tan(eBs)-\gamma^{1}\gamma^{2}\gamma^{\mu}\tan(eBr)-\gamma^{\mu}\tan(eBr)\tan(eBs) \right] \gamma_{\nu}\\
				&=\frac{i\sigma^{\mu\alpha}\tilde{q}_{\alpha}}{2m}\{-8m^{2}\left[1-\tan(eBr)\tan(eBs)\right]\} \\
				&+4m \slashed{q}\gamma^{\mu}\left[1-\tan(eBr)\tan(eBs)\right]+4m \slashed{\hat{q}}\gamma^{\mu}\left[1-\tan(eBr)\tan(eBs)\right]\\
				& + 2m\tilde{q}_{\alpha}(g^{2\alpha}\gamma^{\mu}\gamma^{1}-g^{1\alpha}\gamma^{\mu}\gamma^{2}+i \sigma^{\mu\alpha}\gamma^{1}\gamma^{2})\left[\tan(eBr)+\tan(eBs)\right],
			\end{aligned}
		\end{equation}
		the analysis is in the same way but the longitudinal directions will not contribute an observable AMM when a magnetic field along $z$ axis, so we omit.

	%\end{CJK*}
\end{document}